\documentclass{INTERSPEECH2023}
\usepackage{svg}
\usepackage{float}
\usepackage{wrapfig}
\usepackage{array}
\usepackage{caption}
\usepackage{amsmath}
\usepackage{booktabs}
\usepackage[export]{adjustbox}
\usepackage{graphicx}
\usepackage{hyperref}
\usepackage{multirow}
\usepackage{graphicx}
\usepackage{tabularx}
\usepackage{balance}
\hypersetup{
    colorlinks=true,
    linkcolor=black,
    urlcolor=blue,
    pdftitle={},
    pdfpagemode=FullScreen,
    }
\interspeechcameraready

\title{On the Efficacy and Noise-Robustness of Jointly Learned Speech Emotion and Automatic Speech Recognition}
\name{Lokesh Bansal$^{1,2}$\sthanks{\ \ \ This work was carried out during Bansal's internship at Uniphore Software Systems. He would like to thank Dr. Rajasekhar Bangaru for providing valuable suggestions.}\ \ , S. Pavankumar Dubagunta$^1$, Malolan Chetlur$^1$, \\ \vspace{0.2ex} Pushpak Jagtap$^2$ \textup{and} Aravind Ganapathiraju$^1$}
\address{
  $^1$Uniphore Software Systems, Bengaluru, India \\ %\hspace{5ex}
  $^2$Robert Bosch Centre for Cyber-Physical Systems, Indian Institute of Science, Bengaluru, India}
\email{lokeshbansal@iisc.ac.in, pavankumard@uniphore.com} %, malolan.chetlur@uniphore.com, pushpak@iisc.ac.in, aravindganapathiraju@uniphore.com}

\begin{document}

\maketitle
 
\begin{abstract}
% In this paper, we propose a multitasking architecture that leverages both speech-emotion recognition (SER) and automatic speech recognition (ASR) tasks by combining acoustic and linguistic features. Our architecture improves the accuracy of emotion recognition while also reducing the word error rate (WER) and character error rate (CER) in the speech recognition task. Our proposed architecture uses a shared acoustic model, allowing for the efficient use of shared information between tasks. We show that our approach outperforms single-task models on both emotion recognition and speech recognition tasks. Our experiments demonstrate that the proposed architecture achieves a x improvement in emotion recognition accuracy and a y reduction in WER and CER compared to the single-task baseline models. \textcolor{red}{what is x and y?} These results show the potential of multitasking architectures for improving the accuracy of speech emotion recognition and automatic speech recognition. 
New-age conversational agent systems perform both speech-emotion recognition (SER) and automatic speech recognition (ASR) using two separate and often independent approaches for real-world applications in noisy environments. In this paper, we investigate a joint ASR-SER multitask learning approach in a low-resource setting and show that improvements are observed not only in SER but also in ASR.
% both tasks can leverage each other through sharing of their model parameters.
We also investigate the robustness of such jointly trained models to the presence of background noise, babble, and music.
Experimental results on the IEMOCAP dataset show that joint learning can improve ASR word error rate (WER) and SER classification accuracy by $10.7\%$ and $2.3\%$ respectively in clean scenarios.
% Experimental results show that in terms of ASR, the WER improves by 10.7\%, and the SER accuracy improves by 2.3\% in a noise-free setting.
In noisy scenarios, results on data augmented with MUSAN show that the joint approach outperforms the independent ASR and SER approaches across many noisy conditions. %, except the ASR performance in the presence of background speech.
% We found that modeling noisy data yields improved performance over modeling clean data in most test scenarios, notably except for clean test conditions.
Overall, the joint ASR-SER approach yielded more noise-resistant models than the independent ASR and SER approaches.

\end{abstract}

\section{Introduction}
% ASR is a technology used to recognize and transcribe human speech, and SER is used to recognize and interpret emotions conveyed in speech. ASR has applications in virtual assistants, speech-to-text services, and voice-enabled devices, while SER has applications in areas such as mental health diagnosis, market research, and customer service. Now-a-days, modern conversational agents are equipped with emotional awareness, using both ASR and SER to provide better responses to customers.
Nowadays, modern AI-based conversational agents employing ASR are also equipped with emotional awareness using SER to improve the overall user experience.
These agents are indispensable in real-world applications such as call centers, automotive voice assistants, and voice-enabled home automation systems.
For instance, in call centers, conversational AI agents that automatically transcribe the customers' requests using ASR also detect their emotions using SER to ensure customer satisfaction. Therefore, it is essential to build reliable systems that can perform ASR and SER jointly to simplify the computational requirements in scenarios where both tasks are necessary.

However, integrating ASR and SER presents significant challenges, since (i) their underlying approaches have traditionally been developed separately, (ii) corpora that contain both transcriptions and emotion annotations are scarcely available, limiting their performance in joint-use scenarios. % Despite the considerable scope for improving the accuracy and robustness of ASR and SER through joint training,
There has been limited research in this area, with only a few studies \cite{cai2021, li2022}.
% focusing primarily on SER without any emphasis on the ASR performance and noise robustness
The authors of \cite{cai2021} demonstrated that multitasking ASR and SER jointly during training can enhance SER performance. In \cite{li2022}, the authors showed an improvement in SER performance by fusing acoustic embeddings with linguistic embeddings generated from the ASR output and jointly training the ASR and SER end-to-end model. In both works, ASR was considered as an auxiliary task to SER, i.e. there is no clear emphasis on the ASR performance. Moreover, the noise robustness of these models has not been studied. Operating in a real-world setting requires the models to be robust to background conditions, most typically noise, music, and babble.
% While joint integration of ASR and SER is challenging, noise robustness in ASR and SER models is also crucial for AI-based conversational agents operating in noisy environments. as they are susceptible to various types of background conditions like babble, music and noise.
% Therefore, it is important to investigate and improve the noise robustness of ASR and SER models. 
While several approaches exist that improve the robustness of individual ASR and SER systems in noisy environments, the noise robustness of joint modeling has not been studied. % there has been limited research on jointly enhancing the performance of both tasks in such noisy situations.
% In this context, joint modeling of ASR and SER by explicitly accounting for noise variability during training can improve noise robustness.
% In this context, explicitly including multiple background conditions during training can improve noise robustness.

% To the best of our knowledge, this paper is the first to demonstrate the efficacy and noise resilience of co-learning ASR and SER.

% Building upon the above prior work,
With this motivation, our research presents two main contributions. Firstly, we utilize a multitask joint ASR-SER learning architecture~\cite{li2022}, with both of them as primary tasks, with a minor modification in the fusion strategy of acoustic and linguistic information in addition to the existing fusion approach in \cite{li2022}. This approach overcomes the problem of data scarcity by utilizing pre-trained models, specifically wav2vec2.
% ,
% i.e. different from 
% which the existing literature has not focused on \cite{cai2021, li2022}.
Furthermore, we investigate and present our results on noise robustness analysis carried out to assess the resilience of the multitasking approach to various typical noise kinds, in comparison to the single-tasking baseline approaches.
% We also highlight the use of residual connections with self-attention layers to improve the multitasking architecture, 
% We also investigated improved methods to combine the acoustic and linguistic information for the SER task,
%, especially in low-resource situations.
% In addition to this, we proposed an Adaptive Scaling Loss Function for joint training of ASR and SER tasks.
% TODO: Motivate why we are doing this.
% This loss function utilizes an adaptive scaling factor that adjusts based on the rate of change of the individual loss functions, allowing for dynamic adjustment of the contribution of each task to the final loss based on their relative rates of convergence.
% Towards this, we investigated dynamically combining the ASR and SER losses for improved joint modeling.

The rest of the paper is organized as follows: related work is introduced in Sec.~\ref{sec:related_work}, the multitask-based joint learning approach is described in Sec.~3, the experimental settings and results are presented in Sec.~4, analysis is presented in Sec.~\ref{sec:analysis} and the conclusions are presented in Sec.~6.

\section{Related Work}
\label{sec:related_work}

\begin{figure*}[!htp]
  \centering
  \includegraphics[width=1\textwidth]{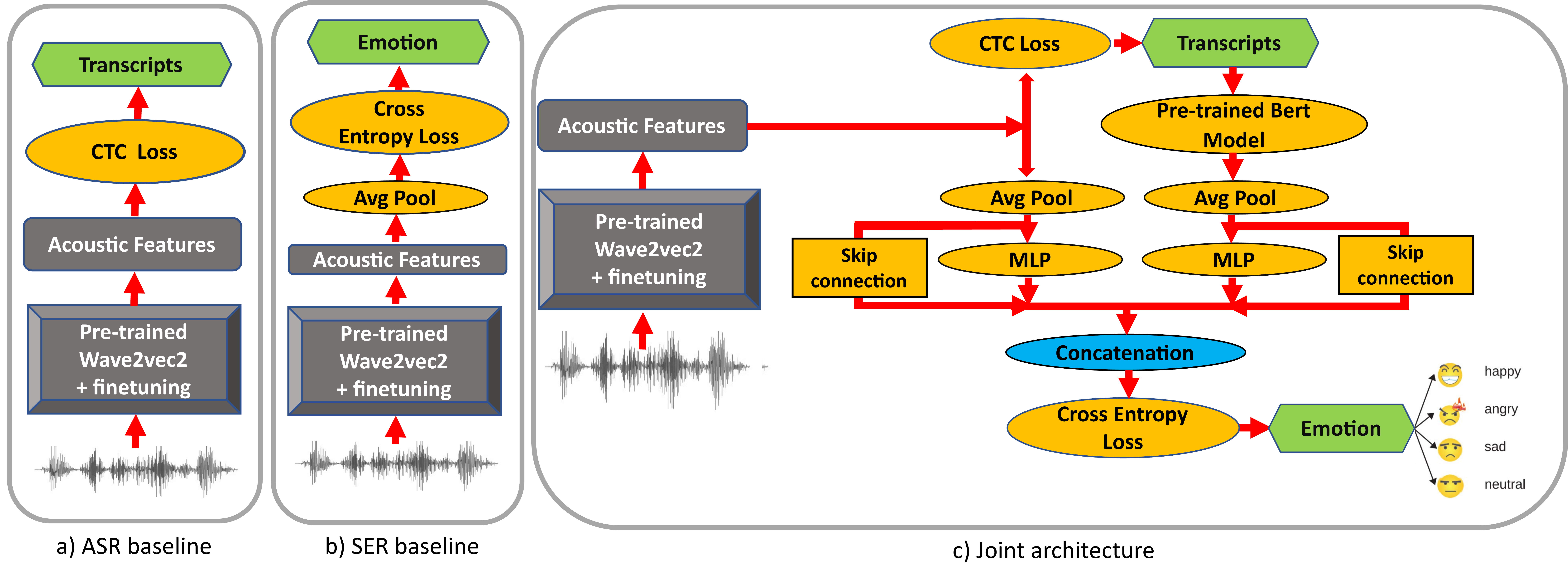}
  \vspace{-3ex}
  \caption{Architectures.}
  \label{fig:Proposed_architecture}
\end{figure*}

According to the literature \cite{benzeguiba2007}, ASR can benefit from modeling variability among data, such as speaker-related (e.g., i-vectors, x-vectors) and channel-related information (through multi-condition training). Several studies have investigated multitasking with various speech tasks, including age, gender, and speaker verification. In \cite{chen2015}, improvements were observed in both ASR and text-dependent speaker verification when performed together. Multitasking with tasks similar to ASR, such as phone label, phone context, and state context prediction, can also improve ASR performance, as demonstrated in \cite{Seltzer2013} and \cite{Arora2017}.
Thus, one can expect that modeling emotion variability by multitasking with emotion recognition can benefit ASR.
Correspondingly, for SER, multi-task learning models have been introduced to predict gender, age, accent, and emotion in a joint manner, and these models demonstrated an enhanced SER performance \cite{Anish2020, Siddique2019, zhao2018, kim2017}. 
As mentioned above, although ASR and SER have been multitasked with other tasks separately, there is also research that combines ASR and SER tasks together which has demonstrated to improve the performance of SER task alone \cite{cai2021, li2022, Santoso2021}, but the improvements on the ASR side have not been systematically evaluated. 

In addition to multitasking ASR and SER using acoustic embeddings, recent studies have emphasized the fusion of acoustic and linguistic embeddings to improve SER performance. Linguistic embeddings offer additional information about spoken words, leading to better emotion recognition accuracy. Combining both acoustic and linguistic embeddings has been demonstrated to enhance SER classifier performance in comparison to using acoustic embeddings alone. Numerous fusion methods, including the Gated Bidirectional Alignment Network (GBAN) proposed by \cite{Liu2020}, the deep dual recurrent encoder model proposed by \cite{Lee2020}, and the fusion techniques proposed by Sebastian et al. \cite{Sebastian2019}, have displayed the potential for improving emotion recognition using multiple modalities. Following the previous fusion methods, we present a simple modification to the architecture by incorporating skip connections in conjunction with feed-forward layers. This revised approach entails processing the two embeddings separately using the skip connection-based feed-forward layer before concatenating them.

As advancements have been made in ASR and SER for noise-free environments, there have also been independent studies on ASR and SER in noisy environments. Several techniques were proposed to improve noise robustness in ASR, such as the interactive feature fusion network (IFF-Net) \cite{Yuchen2022}, the multitask-based method with dual-channel data augmentation \cite{Duo2021}, and an online speech distortion module with an encoder \cite{Mirco2020}, to name a recent few. On the other hand, techniques like the adversarial joint training framework with the self-attention mechanism \cite{Lujun2021} and data augmentation techniques, including overlaying background noise and loudness variations, have been proposed to improve noise robustness in SER models \cite{Egor2018}. Additionally, a feature selection technique based on the noise robustness of individual low-level descriptors in noisy conditions has been proposed to improve SER performance without using any compensation methods or a noisy training set \cite{Leem}.

Since we have observed improvements in ASR-SER models through joint training, it is important to investigate the effect of joint ASR-SER training on noise robustness in real-world settings where ASR-SER applications are typically used. In such scenarios, it is critical to ensure that ASR-SER joint models are able to perform accurately in noisy environments.

\section{Approach}

% \usepackage{booktabs}
% \usepackage{multirow}
% \usepackage{graphicx}
\begin{table*}
\caption{Results of 10-fold cross-validation on IEMOCAP dataset for all the models trained on Clean and Noise data and evaluated at seven test scenarios. \emph{Rel Imp} refers to relative improvement (\%).}
\resizebox{\textwidth}{!}{%
\begin{tabular}{@{}|l|cccccc|cccccc|@{}}
\toprule
\multicolumn{1}{|c|}{\multirow{2}{*}{\textbf{\begin{tabular}[c]{@{}c@{}}10 Fold Cross \\ Results\end{tabular}}}} &
  \multicolumn{6}{c|}{\textbf{ASR Performance: Metric Word Error 
  Rate (WER)}} &
  \multicolumn{6}{c|}{\textbf{SER Performance: Metric - Accuracy}} \\ \cmidrule(l){2-13} 
\multicolumn{1}{|c|}{} &
  \multicolumn{3}{c|}{\textbf{Trained on Clean}} &
  \multicolumn{3}{c|}{\textbf{Trained on   Noise}} &
  \multicolumn{3}{c|}{\textbf{Trained on Clean}} &
  \multicolumn{3}{c|}{\textbf{Trained on   Noise}} \\ \midrule
\multicolumn{1}{|c|}{\textbf{Test Dataset}} &
  \multicolumn{1}{l}{\textbf{Baseline}} &
  \multicolumn{1}{l}{\textbf{Joint}} &
  \multicolumn{1}{l|}{\textbf{Rel Imp}} &
  \multicolumn{1}{l}{\textbf{Baseline}} &
  \multicolumn{1}{l}{\textbf{Joint}} &
  \multicolumn{1}{l|}{\textbf{Rel Imp}} &
  \multicolumn{1}{l}{\textbf{Baseline}} &
  \multicolumn{1}{l}{\textbf{Joint}} &
  \multicolumn{1}{l|}{\textbf{Rel Imp}} &
  \multicolumn{1}{l}{\textbf{Baseline}} &
  \multicolumn{1}{l}{\textbf{Joint}} &
  \multicolumn{1}{l|}{\textbf{Rel Imp}} \\ \midrule
\textbf{Clean} &
  16.8 &
  15.0 &
  \multicolumn{1}{c|}{10.7} &
  19.8 &
  15.6 &
  21.2 &
  71.9 &
  74.2 &
  \multicolumn{1}{c|}{2.3} &
  70.2 &
  71.7 &
  1.5 \\
\textbf{SNR 15 Noise} &
  22.2 &
  19.3 &
  \multicolumn{1}{c|}{13.1} &
  19.8 &
  18.4 &
  7.1 &
  69.8 &
  71.8 &
  \multicolumn{1}{c|}{2.0} &
  69.0 &
  70.6 &
  1.6 \\
\textbf{SNR 5 Noise} &
  33.6 &
  29.3 &
  \multicolumn{1}{c|}{12.8} &
  26.0 &
  25.8 &
  0.8 &
  62.0 &
  66.8 &
  \multicolumn{1}{c|}{4.8} &
  66.1 &
  68.7 &
  2.6 \\
\textbf{SNR 15 Music} &
  21.7 &
  19.6 &
  \multicolumn{1}{c|}{9.9} &
  19.1 &
  18.3 &
  4.3 &
  68.8 &
  72.1 &
  \multicolumn{1}{c|}{3.3} &
  68.7 &
  71.1 &
  2.4 \\
\textbf{SNR 5 Music} &
  37.4 &
  35.8 &
  \multicolumn{1}{c|}{4.3} &
  27.9 &
  29.2 &
  -4.6 &
  60.8 &
  64.2 &
  \multicolumn{1}{c|}{3.4} &
  65.1 &
  68.4 &
  3.3 \\
\textbf{SNR 15 Speech} &
  36.9 &
  42.0 &
  \multicolumn{1}{c|}{-13.8} &
  20.6 &
  22.4 &
  -8.6 &
  66.7 &
  67.6 &
  \multicolumn{1}{c|}{0.9} &
  67.4 &
  70.0 &
  2.6 \\
\textbf{SNR 5 Speech} &
  74.6 &
  86.4 &
  \multicolumn{1}{c|}{-15.8} &
  36.3 &
  51.3 &
  -41.3 &
  53.3 &
  52.4 &
  \multicolumn{1}{c|}{-0.9} &
  58.1 &
  62.3 &
  4.2 \\ \bottomrule
\end{tabular}%
}
\label{table:iemocap}
\end{table*}

We begin with the description of the baseline models for ASR and SER tasks, which will serve as a point of comparison for the joint architecture. Then, we will present the joint architecture for a multitask-based joint learning approach. Figure \ref{fig:Proposed_architecture} shows the overall architecture of the joint multitasking model along with the ASR baseline and SER baseline models. 
In our experiments, we employed the SpeechBrain toolkit \cite{SpeechBrain} for both the joint architecture and baselines, since its recipes are considered state-of-the-art. We used the recipes provided by the toolkit as-is for the baselines, which included the ASR and SER baselines. To implement the joint-training approach, we introduced modifications to the same recipe to attain the desired joint architecture. More comprehensive explanations of these modifications will be provided in Sec.~\ref{subsec:joint_architecture}.

\subsection{Baseline for ASR and SER}

 The ASR baseline involves performing data augmentation using Time Domain Spec Augment with three different speed values 95, 100, and 105, followed by acoustic feature extraction using the Hugging Face wav2vec2-large-960h-lv60-self wav2vec2 model \cite{Wav2vec2}. The wav2vec2 encoder was fine-tuned while keeping the convolutional part of the model frozen. The training (fine-tuning) was performed using connectionist temporal classification (CTC) loss \cite{CTC} and the decoded transcripts were obtained using the corresponding decoding algorithm.
The SER baseline involves acoustic feature extraction using the same wav2vec2 encoder, with its fine-tuning with the convolutional part of the encoder frozen. It is followed by an average pooling layer and an output MLP for emotion classification into anger, happiness, sadness, and neutral categories.
\subsection{Joint architecture}
\label{subsec:joint_architecture}
The joint multitasking architecture aims to improve the performance of both ASR and SER tasks. Firstly, the ASR baseline approach is followed, which involves data augmentation and fine-tuning a pre-trained wav2vec2 model as a shared acoustic feature extractor for both tasks, while freezing the convolutional part of the model. These acoustic feature embeddings are then processed through two parallel and interlinked channels, one for ASR and the other for SER. The ASR channel utilizes a CTC-based decoder \cite{CTC} to decode the acoustic features and obtain speech transcriptions. The transcriptions contain emotional content as specified in lexical dictionaries \cite{hissell, Warriner, Hutto, Mohammad}. To leverage this information, linguistic embeddings are extracted from the ASR transcriptions using a pre-trained BERT encoder, which has a greater representation capability than other types of embeddings due to its context-aware representations, resulting in more robust and accurate embeddings.
Then these BERT-based linguistic embeddings are used in a fusion network for feature-level fusion with the acoustic embeddings. The fusion approach we utilized processes linguistic and acoustic embeddings separately through multi-layer perceptron (MLP). To enhance the robustness, and information flow of these layers, we incorporate skip connections \cite{resnet} across them that results in more accurate and reliable embeddings. We
%transformer encoder with 4 heads each for self-attention, 
concatenate the processed embeddings to obtain the final embeddings for SER classification. Finally, a cross-entropy loss-based emotion classifier is used for SER classification.
For the training of the joint architecture, we used the following multitasking loss function: 

\begin{equation}
  \label{eq:joint_loss}
  \mathcal{L}_{joint} = \alpha \mathcal{L}_{SER} + (1 - \alpha) \mathcal{L}_{ASR}
\end{equation}

Where $\mathcal{L}_{SER}$ is the SER cross-entropy loss, $\mathcal{L}_{ASR}$ is the ASR CTC loss and $\alpha$ is a scaling factor that determines the contribution of the SER and ASR components. We used $\alpha = 0.1$ that gave the best results (analysis not shown) for both the tasks as primary tasks.

\section{Experiments and Results}

\subsection{Dataset preparation and Experimental setup}

For our study, we utilized the IEMOCAP dataset \cite{IEMOCAP}, which consists of approximately 12 hours of data from ten actors in five dyadic sessions. The dataset contains 10,039 turns or utterances, sampled at 16 kHz. We used Speechbrain toolkit \cite{SpeechBrain} for experimentation - specifically the `IEM4' categories which include happy, sad, angry, and neutral emotions. In total, we used 5500 utterances in the IEM4 categories for training and evaluation.
Further, to generate noise in the data we used MUSAN (Music, Speech, and Noise) dataset \cite{MUSAN}. The MUSAN dataset is a large-scale corpus of various types of audio signals, such as music, speech, and noise. The dataset includes more than 8,000 audio files, each 10 seconds long, sampled at 16 kHz. The audio files are categorized into three groups, namely music, speech, and noise, and further sub-categorized based on various properties such as genre, language, and type of noise.

For training the baseline and joint models, we used two types of data: clean data and noisy data. The IEMOCAP IEM4 data was used as the clean set, while the noisy data was created by superimposing the IEM4 data with audio files from the MUSAN dataset,
% . The MUSAN data allowed to the addition of random background noise to the IEMOCAP data,
resulting in noisy files with speech-to-noise ratio (SNR) values ranging from 5 to 35 dB and noise types including music, background speech (babble), and background noise. Using these two types of datasets, we trained three types of models (joint architecture, SER, and ASR baseline), resulting in six types of trained models.
% (clean-trained SER baseline, noise-trained SER baseline, clean-trained ASR baseline, noise-trained ASR baseline, the clean-trained proposed model and noise-trained proposed model)
To evaluate these models, we prepared seven test scenarios with 15 and 5 dB SNR levels across noise types.
% : Clean, SNR 15 Music, SNR 5 Music, SNR 15 Speech, SNR 5 Speech, SNR 15 Noise, and SNR 5 Noise.
The joint and baseline methods were evaluated using 10-fold cross-validation. Following the standard practice, one speaker was excluded for testing in each fold, while the remaining nine speakers were used for training. We applied this evaluation approach to the IEMOCAP dataset, which consists of 10 speakers. In each fold of cross-validation, we held out one speaker at a time for testing and utilized the remaining nine speakers for training. During this process, we collected and stored the predictions for emotion and transcription. This hold-and-training procedure was repeated for all ten folds, and finally, the overall performance was assessed by analyzing the predictions. 

% The evaluation involves 10-fold cross-validation over the IEMOCAP dataset for the ASR baseline, SER baseline, and proposed architecture. 

\begin{figure*}[!htp]
  \centering
  \includegraphics[width=1\textwidth]{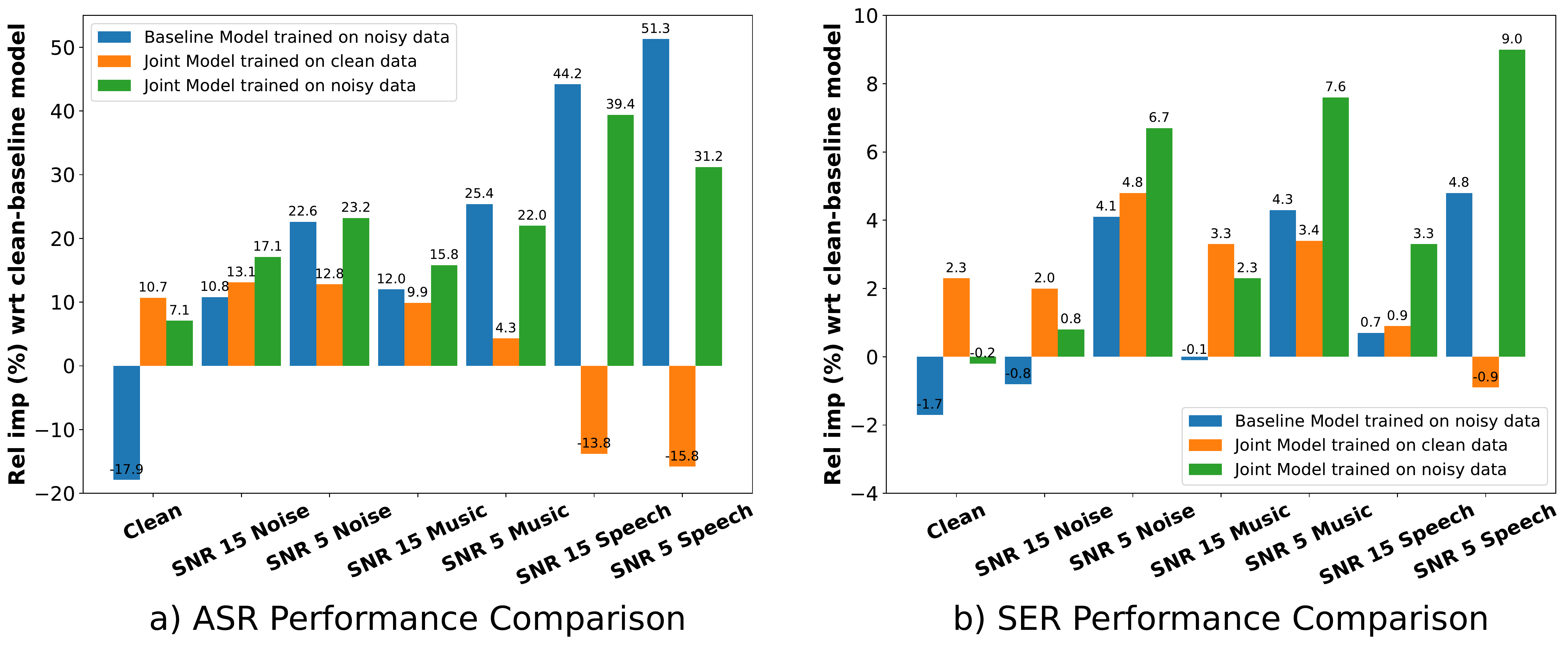}
  \caption{Relative comparison among the baseline models and joint architecture trained with the two training strategies: Clean and Noisy dataset.}  \vspace{-3ex}
  \label{fig:Compare_ASR_SER_performance_of_proposed_model_trained_on_noisy_vs_clean_data}
\end{figure*}

% \label{table:iemocap}
\subsection{Results}
Table \ref{table:iemocap} showcases the outcomes of the aforementioned experiments.
% 10-fold cross-validation results for ASR and SER performance of the proposed and baseline architectures in the seven test case scenarios as mentioned in the above section.
The ASR performance is measured in word error rate (WER) (the lower the better), while the SER performance is determined by classification accuracy (the higher the better). The relative performance improvement was calculated by comparing the clean-trained joint model to the clean-trained baseline models and the noise-trained joint model to the noise-trained baseline models. %; however, there were a few instances where the baseline system performed better than the proposed architecture, specifically for ASR in the scenario of speech that was corrupted by background speech.
% In most cases, the joint model outperformed the baseline models, barring a few exceptions.

% To begin with, we examine the noise-free scenario where both the proposed and baseline models were trained and tested on a clean dataset.
The results of training testing performed on clean data show that the joint model significantly outperformed the baseline models for both ASR and SER. In terms of ASR, there was a 10.7\% improvement in the WER, from 16.8 to 15, and a 2.3\% improvement in SER accuracy, from 71.9\% to 74.2\%.

\section{Analysis}
\label{sec:analysis}
In this section, we examine the variation in performance between the baseline and joint models that were trained on clean and noisy data.  We use the baseline models trained on clean data as the reference point for the other three types of models, which include the noise-trained baseline models, the clean-trained joint model, and the noise-trained joint model.  Fig.\ref{fig:Compare_ASR_SER_performance_of_proposed_model_trained_on_noisy_vs_clean_data}(a) and (b) show these comparisons.
% Plot in Fig.\ref{fig:Compare_ASR_SER_performance_of_proposed_model_trained_on_noisy_vs_clean_data}(a) illustrates the relative improvements in WER. Plot in Fig.\ref{fig:Compare_ASR_SER_performance_of_proposed_model_trained_on_noisy_vs_clean_data}(b) illustrates the relative improvements in SER accuracy.

% Let's start by examining the ASR task performance. 
First, we evaluate the impact of noise inclusion in the training data for the seven test scenarios. %The plot in 
Fig.\ref{fig:Compare_ASR_SER_performance_of_proposed_model_trained_on_noisy_vs_clean_data}(a) shows that, irrespective of the baseline or joint model, training on a noisy dataset yields better performance in all the ASR test cases, except for clean test speech. 

% \begin{figure}[H]
%   \centering
%   \includegraphics[width=1\columnwidth]{Compare_ASR_performance_of_proposed_model_trained_on_noisy_vs_clean_data.pdf}
%   \caption{Compare ASR performance of the baseline and proposed model trained on noisy vs clean datasets}
%   \label{fig:Relative_ASR_performance_various_models}
% \end{figure}

Furthermore, when we compare the baseline model to the joint model, regardless of clean or noisy training data, the joint model performs better when there is music and background noise mixed with the data, across all levels of SNR except for scenario SNR 5 Music, where there is a slight improvement in baseline in comparison to the joint model. In Table \ref{table:iemocap}, we can observe a significant relative improvement of the clean-trained joint model over the clean-trained baseline model in the case of noise and music in the background, with a huge improvement across all SNR levels, and the highest improvement of 13.1\% for the SNR 15 noise case. Similarly, for the comparison between noise-trained joint and noise-trained baseline models, we can see a significant improvement for the clean test as well as the noise and music-contaminated test set across all SNR levels, with a remarkable improvement of 21.2\% for the clean data and significant improvement across other mentioned data sets. Hence, it can be concluded that the joint model always outperforms the baseline model for all SNR levels for music and noise speech contamination scenarios. However, in the case of speech in the background, the baseline model performs better than the joint model, regardless of training type and SNR level. This observation requires further investigation. 
% If the model cannot differentiate between the primary speech and the background speech, the ASR task's performance will suffer. 
% \begin{figure}[H]
%   \centering
%   \includegraphics[width=1\columnwidth]{Compare_SER_performance_of_proposed_model_trained_on_noisy_vs_clean_data.pdf}
%   \caption{Compare SER performance of the baseline and proposed model trained on noisy and clean datasets}
%   \label{fig:Relative_SER_performance_various_models}
% \end{figure}

In the case of SER, we first focus on the impact of a clean versus noisy training approach. As shown in Figure \ref{fig:Compare_ASR_SER_performance_of_proposed_model_trained_on_noisy_vs_clean_data}(b), with the exception of the clean test and SNR 15 noise and SNR 15 Music scenarios, the noisy training approach outperforms the clean training approach for both the joint and baseline  models. The joint model trained with noisy data exhibits exceptional performance, particularly in the babble scenario and even in highly contaminated scenarios such as the SNR 5 level across all noise types. The most notable improvement is observed in the SNR 5 Noise case, where it outperforms the clean-trained joint model by 9.9\% and the noise-trained baseline model by 5.2\%. Additionally, the comparison between the noise-trained SER baseline and the clean-trained SER baseline indicates a trend similar to that found in the joint model's noise versus clean case. This suggests that noise training enhances the noise robustness of the model, regardless of the baseline or joint architecture, particularly at low SNR levels and babble scenarios. However, in other cases, the clean-trained models perform better than the noise-trained models.

% After examining the importance of noise training for the SER task,
Finally, we analyze the performance of the joint architecture versus the baseline architecture for both clean and noise-trained scenarios. As shown in the plot, the joint model consistently outperforms the baseline model with significant margins (except for the clean training at SNR 5 speech). This observation indicates that the joint architecture is more robust than the baseline architecture for all types of noise at all SNR levels.
Overall for SER, the joint architecture demonstrates better noise robustness than the baseline model regardless of the training strategy.

\section{Conclusions}
The results conclude that joint modeling of ASR-SER improves both tasks in clean conditions over the single-tasking baselines.
In the case of ASR, training on a noisy dataset results in better performance for all test scenarios except clean speech, regardless of whether it is a baseline or joint model. Additionally, the joint model consistently exhibits better performance than the baseline model in most of the test scenarios, except for the babble scenario and the SNR 5 Music scenario (when noise training is used). For the babble scenario, regardless of the training approach, ASR performance degrades at both 15 and 5 SNR, with higher degradation at 5 SNR. Further, in the case of SER, noise training enhances the noise robustness of the model in low SNR levels and babble, and the joint architecture consistently outperforms the baseline architecture for all types of noise and SNR levels. 

Finally, we found that joint architectures are generally more noise robust than baseline architectures, but there are still some scenarios where the baseline outperforms the joint model, particularly for ASR tasks having babble test scenarios and Music with lower SNR. Further research could investigate why this is the case and develop strategies to improve joint models' performance in these scenarios.
% Therefore, it can be concluded that, the noise in the training data affects the performance of the ASR and SER models differently depending on the type of noise and SNR level in the test dataset. Further, the proposed model is very robust toward noise over the baseline models.

\bibliographystyle{IEEEtran}
\balance
\bibliography{main}

\begin{thebibliography}{10}
\providecommand{\url}[1]{#1}
\csname url@samestyle\endcsname
\providecommand{\newblock}{\relax}
\providecommand{\bibinfo}[2]{#2}
\providecommand{\BIBentrySTDinterwordspacing}{\spaceskip=0pt\relax}
\providecommand{\BIBentryALTinterwordstretchfactor}{4}
\providecommand{\BIBentryALTinterwordspacing}{\spaceskip=\fontdimen2\font plus
\BIBentryALTinterwordstretchfactor\fontdimen3\font minus
  \fontdimen4\font\relax}
\providecommand{\BIBforeignlanguage}[2]{{%
\expandafter\ifx\csname l@#1\endcsname\relax
\typeout{** WARNING: IEEEtran.bst: No hyphenation pattern has been}%
\typeout{** loaded for the language `#1'. Using the pattern for}%
\typeout{** the default language instead.}%
\else
\language=\csname l@#1\endcsname
\fi
#2}}
\providecommand{\BIBdecl}{\relax}
\BIBdecl

\bibitem{cai2021}
X.~Cai, J.~Yuan, R.~Zheng, L.~Huang, and K.~Church, ``Speech emotion
  recognition with multi-task learning.'' in \emph{Proceedings of Interspeech},
  2021, pp. 4508--4512.

\bibitem{li2022}
Y.~Li, P.~Bell, and C.~Lai, ``Fusing {ASR} outputs in joint training for speech
  emotion recognition,'' in \emph{Proceedings of ICASSP}.\hskip 1em plus 0.5em
  minus 0.4em\relax IEEE, 2022, pp. 7362--7366.

\bibitem{benzeguiba2007}
M.~Benzeghiba \emph{et~al.}, ``Automatic speech recognition and speech
  variability: A review,'' \emph{Speech communication}, vol.~49, no. 10-11, pp.
  763--786, 2007.

\bibitem{chen2015}
N.~Chen, Y.~Qian, and K.~Yu, ``{Multi-task learning for text-dependent speaker
  verification},'' in \emph{Proceedings of Interspeech}, 2015, pp. 185--189.

\bibitem{Seltzer2013}
M.~L. Seltzer and J.~Droppo, ``Multi-task learning in deep neural networks for
  improved phoneme recognition,'' in \emph{Proceedings of ICASSP}.\hskip 1em
  plus 0.5em minus 0.4em\relax IEEE, 2013, pp. 6965--6969.

\bibitem{Arora2017}
V.~Arora, A.~Lahiri, and H.~Reetz, ``{Phonological Feature Based
  Mispronunciation Detection and Diagnosis Using Multi-Task DNNs and Active
  Learning},'' in \emph{Proceedings of Interspeech}, 2017, pp. 1432--1436.

\bibitem{Anish2020}
A.~Nediyanchath, P.~Paramasivam, and P.~Yenigalla, ``Multi-head attention for
  speech emotion recognition with auxiliary learning of gender recognition,''
  in \emph{ICASSP}.\hskip 1em plus 0.5em minus 0.4em\relax IEEE, 2020, pp.
  7179--7183.

\bibitem{Siddique2019}
S.~Latif \emph{et~al.}, ``Multi-task semi-supervised adversarial autoencoding
  for speech emotion recognition,'' \emph{IEEE Transactions on Affective
  Computing}, vol.~13, no.~2, pp. 992--1004, 2022.

\bibitem{zhao2018}
H.~Zhao, N.~Ye, and R.~Wang, ``Transferring age and gender attributes for
  dimensional emotion prediction from big speech data using hierarchical deep
  learning,'' in \emph{2018 IEEE 4th International Conference on Big Data
  Security on Cloud (BigDataSecurity), IEEE International Conference on High
  Performance and Smart Computing, (HPSC) and IEEE International Conference on
  Intelligent Data and Security (IDS)}, 2018, pp. 20--24.

\bibitem{kim2017}
J.~Kim, G.~Englebienne, K.~P. Truong, and V.~Evers, ``{Towards Speech Emotion
  Recognition “in the Wild” Using Aggregated Corpora and Deep Multi-Task
  Learning},'' in \emph{Proceedings of Interspeech}, 2017, pp. 1113--1117.

\bibitem{Santoso2021}
J.~Santoso \emph{et~al.}, ``Speech emotion recognition based on attention
  weight correction using word-level confidence measure.'' in \emph{Proceedings
  of Interspeech}, 2021, pp. 1947--1951.

\bibitem{Liu2020}
P.~Liu, K.~Li, and H.~Meng, ``{Group Gated Fusion on Attention-Based
  Bidirectional Alignment for Multimodal Emotion Recognition},'' in
  \emph{Proceedings of Interspeech}, 2020, pp. 379--383.

\bibitem{Lee2020}
L.~Sun, B.~Liu, J.~Tao, and Z.~Lian, ``Multimodal cross-and self-attention
  network for speech emotion recognition,'' in \emph{Proceedings of
  ICASSP}.\hskip 1em plus 0.5em minus 0.4em\relax IEEE, 2021, pp. 4275--4279.

\bibitem{Sebastian2019}
J.~Sebastian, P.~Pierucci \emph{et~al.}, ``Fusion techniques for
  utterance-level emotion recognition combining speech and transcripts.'' in
  \emph{Proceedings of Interspeech}, 2019, pp. 51--55.

\bibitem{Yuchen2022}
Y.~Hu, N.~Hou, C.~Chen, and E.~S. Chng, ``Interactive feature fusion for
  end-to-end noise-robust speech recognition,'' in \emph{Proceedings of
  ICASSP}.\hskip 1em plus 0.5em minus 0.4em\relax IEEE, 2022, pp. 6292--6296.

\bibitem{Duo2021}
D.~Ma, N.~Hou, H.~Xu, E.~S. Chng \emph{et~al.}, ``Multitask-based joint
  learning approach to robust asr for radio communication speech,'' in
  \emph{Proceedings of Asia-Pacific Signal and Information Processing
  Association Annual Summit and Conference (APSIPA ASC)}.\hskip 1em plus 0.5em
  minus 0.4em\relax IEEE, 2021, pp. 497--502.

\bibitem{Mirco2020}
Ravanelli \emph{et~al.}, ``Multi-task self-supervised learning for robust
  speech recognition,'' in \emph{Proceedings of ICASSP}.\hskip 1em plus 0.5em
  minus 0.4em\relax IEEE, 2020, pp. 6989--6993.

\bibitem{Lujun2021}
L.~Li, Y.~Kang, Y.~Shi, L.~K{\"u}rzinger, T.~Watzel, and G.~Rigoll,
  ``Adversarial joint training with self-attention mechanism for robust
  end-to-end speech recognition,'' \emph{EURASIP Journal on Audio, Speech, and
  Music Processing}, vol. 2021, pp. 1--16, 2021.

\bibitem{Egor2018}
E.~Lakomkin, M.~A. Zamani, C.~Weber, S.~Magg, and S.~Wermter, ``On the
  robustness of speech emotion recognition for human-robot interaction with
  deep neural networks,'' in \emph{Proceedings of IEEE/RSJ International
  Conference on Intelligent Robots and Systems (IROS)}.\hskip 1em plus 0.5em
  minus 0.4em\relax IEEE, 2018, pp. 854--860.

\bibitem{Leem}
S.-G. Leem, D.~Fulford, J.-P. Onnela, D.~Gard, and C.~Busso, ``Not all features
  are equal: Selection of robust features for speech emotion recognition in
  noisy environments,'' in \emph{Proceedings of ICASSP}.\hskip 1em plus 0.5em
  minus 0.4em\relax IEEE, 2022, pp. 6447--6451.

\bibitem{SpeechBrain}
M.~Ravanelli, T.~Parcollet, P.~Plantinga, A.~Rouhe, S.~Cornell, L.~Lugosch,
  C.~Subakan, N.~Dawalatabad, A.~Heba, J.~Zhong \emph{et~al.}, ``Speechbrain: A
  general-purpose speech toolkit,'' \emph{arXiv preprint arXiv:2106.04624},
  2021.

\bibitem{Wav2vec2}
A.~Baevski, Y.~Zhou, A.~Mohamed, and M.~Auli, ``wav2vec 2.0: A framework for
  self-supervised learning of speech representations,'' \emph{Advances in
  neural information processing systems}, vol.~33, pp. 12\,449--12\,460, 2020.

\bibitem{CTC}
A.~Graves, S.~Fern{\'a}ndez, F.~Gomez, and J.~Schmidhuber, ``Connectionist
  temporal classification: labelling unsegmented sequence data with recurrent
  neural networks,'' in \emph{Proceedings of the 23rd international conference
  on Machine learning}, 2006, pp. 369--376.

\bibitem{hissell}
C.~Whissell, ``Using the revised dictionary of affect in language to quantify
  the emotional undertones of samples of natural language,''
  \emph{Psychological reports}, vol. 105, no.~2, pp. 509--521, 2009.

\bibitem{Warriner}
A.~B. Warriner, V.~Kuperman, and M.~Brysbaert, ``Norms of valence, arousal, and
  dominance for 13,915 english lemmas,'' \emph{Behavior research methods},
  vol.~45, pp. 1191--1207, 2013.

\bibitem{Hutto}
C.~Hutto and E.~Gilbert, ``Vader: A parsimonious rule-based model for sentiment
  analysis of social media text,'' in \emph{Proceedings of the international
  AAAI conference on web and social media}, vol.~8, no.~1, 2014, pp. 216--225.

\bibitem{Mohammad}
S.~Mohammad, ``Obtaining reliable human ratings of valence, arousal, and
  dominance for 20,000 english words,'' in \emph{Proceedings of the 56th annual
  meeting of the association for computational linguistics (volume 1: Long
  papers)}, 2018, pp. 174--184.

\bibitem{resnet}
K.~He, X.~Zhang, S.~Ren, and J.~Sun, ``Deep residual learning for image
  recognition,'' in \emph{Proccedings of CVPR}, 2016, pp. 770--778.

\bibitem{IEMOCAP}
C.~Busso \emph{et~al.}, ``{IEMOCAP}: Interactive emotional dyadic motion
  capture database,'' \emph{Language resources and evaluation}, vol.~42, pp.
  335--359, 2008.

\bibitem{MUSAN}
D.~Snyder, G.~Chen, and D.~Povey, ``{MUSAN}: {A} {M}usic, {S}peech, and {N}oise
  {C}orpus,'' 2015, arXiv:1510.08484v1.

\end{thebibliography}

% \end{thebibliography}

\end{document}